# Data selection and confounding in the court case of Lucia de Berk

Thomas Colignatus
http://www.dataweb.nl/~cool
August 9 2007

■ **Summary**

The nurse Lucia de Berk was convicted by the Dutch courts as a serial killer with 7 murders and 3 attempts at murder in three hospitals where she worked. The nurse however always professed her innocence and indeed was never observed in such an act of murder. The courts based their decision on circumstantial evidence and upon the use of statistics. In the appeal court, the use of statistical calculations was repealed but the use of "data" and "statistical insights" were not excluded. The trial hinged importantly on the role of statistics and data gathering. It appears that data selection and confounding feature strongly in this case. The notion of "nominal correlation" can be used to highlight those two features. This suggests a mistrial with the conviction of an innocent person.

## Introduction

The Dutch Penal Court (March 24 2003), the Penal Court of Appeal (June 18 2004) and the Supreme Penal Court (March 14 2006) likely conducted a mistrial when they consecutively convicted the nurse Lucia de Berk (Lucia de B.) for 7 murders and 3 attempts at murder in three hospitals where she worked. Lucia de Berk however always professed her innocence and indeed was never observed in such an act of murder. The courts based their decision on circumstantial evidence and upon the use of statistics. In the appeal, the use of statistical calculations was repealed but the use of "data" and "statistical insights" were not excluded. The mistrial thus hinges importantly on the role of statistics and data gathering. Yorker et al. (2006), "Serial Murder by Healthcare Professionals", mention her as one of various cases, but, while the statistics on such serial killings seem doubtful anyway since those occur only rarely, the inclusion of Lucia de Berk in that list uncritically follows juridical outcomes and seems highly premature, since it likely isn't a "case" at all. Buchanan (2007) mentions Lucia de Berk in the context of the Sally Clark trial in the UK, on the sudden infant death syndrome and the abuse of statistics by Sir Roy Meadow, and remarks that "Statistics have the power to trip everyone up".

English sources on the statistics of the case are Meester et al. (2006) with a commentary by Lucy (2007). Other sources are in Dutch only. Derksen (2006) is the authoritative source for judging the issue, reviewing both legal, medical and statistical aspects, and is in clear opposition to the Court, diagnosing a mistrial. Elffers (2002) is the original report by the statistician for the prosecution and the Court. The court case became prominent in the newsmedia while the statistics in the case drew the attention of academic statisticians. Colignatus (2007c) (in Dutch) evaluates the discussion by Dutch statisticians in 2002-2006. The case is complex since one has to take account of all details but once that has been done then some clear lines can be identified. Originally, there were only "incidents", apparently not always defined clearly, but in Elffers (2002) taken as "situations where a patient had to be reanimated, with or without ensuing death". Only the decision by the Court turned some of them into "murder" but in questionable manner. The Court of Appeal assigns only 2 "incidents" as "directly proven murder", though Derksen (2006) explains that the Court uses circumstantial and even wrong "evidence" here. The Court deals with the other 8 "incidents" only in an indirect manner and argues that, since Lucia de Berk was on shift, these must also have been "(attempted) murders by her" too. This might be called the "Dutch Penal Court *chain statistics* verdict". In the original conviction, the Court referred to Elffers (2002) for the statistical significance of the co-occurrence of the "incidences" and Lucia de Berk's roster data. His key conclusion was that this co-occurrence "was no chance". But precisely the method of using statistical significance does not exclude that such co-occurrence remains a matter of chance, and chance only, since precisely the significance level expresses how much chance there still can be.

The mentioned evaluation shows that Elffers made also other mistakes besides the mis-representation of the meaning of hypothesis testing. The evaluation also shows that other statisticians later called attention to those mistakes. For example, the orginal "data" appear to have suffered from data-selection. The "incidents" were counted when they occurred during Lucia de Berk's shifts while other events were discarded when occurring during someone else's shift. Data collection by the prosecution is different from what one would do for statistical purposes. In July 2007, Elffers still doesn't correct his exposed errors and still doesn't repeal his original statement for the Court, turning the issue into one on the integrity of science since a rule is that a researcher should be open for critique and accept the need for correction which such is pointed out.

The following will discuss some main elements in the statistics of the Lucia de Berk court case. It appears that the issues of (1) ***data-selection*** and (2) ***confounding*** featured strongly in this case, while those can be explained in a very accessible manner. Recall that common human decision making does not rely on statistical test procedures. People making a decision consider the data, consider the logic in the situation and arrive at a decision, in ways that are not always clear. It is an added feature of statistics that it also provides the statistical test procedure that allows an expression of the certainty or uncertainty of the decision made. However, when you give a set of data to a statistician then it may be a bit of a pitfall that the focus is narrowed to statistical tests. Looking at data, you should not forget that you still might arrive at some decision without actually doing a statistical test. Such a decision may be weaker than the one based upon a test since you would not be able to specify the amount of certainty or uncertainty of the decision, yet, it would remain a decision. In the following we will use the notion of correlation to clarify the issues of data-selection and confounding in this court case. This notion does not (yet) allow a test but it can clarify issues and contribute to decision making.

Though the notion of correlation is well established in statistics, its application ot nominal data is new. Given this novelty, the discussion below would merit critical consideration. While the method itself is dicussed in Colignatus (2007d) this present paper would be a first application to real data.

The issue of confounding is exemplified in the Simpson paradox that we hence shall discuss first.

## An example of the Simpson paradox

The Simpson paradox is that something may hold for subgroups but not for the total group. A decision based upon subgroups would turn around if one would consider only the total. The reason for such a switch can lie in *confounding*. The following example is taken from Kleinbaum et al. (2003:277). The story is that there are two shops selling both blue and green hats while a customer visits both shops and tries all hats. In each separate shop the green hats fit relatively better, but for both shops combined the blue hats fit relatively better. The dispersion over the two shops is a confounder.

```
TableForm[mat = {{{5, 1}, {8, 2}}, {{2, 8}, {1, 5}}},
 TableHeadings → {{"Shop1", "Shop2"}, {"Green", "Blue"}, {"Fit", "No fit"}}]
```

|       |        | Green | Blue |
|-------|--------|-------|------|
| Shop1 | Fit    | 5     | 8    |
|       | No fit | 1     | 2    |
| Shop2 | Fit    | 2     | 1    |
|       | No fit | 8     | 5    |

When we consider the overall association or correlation between these data (Colignatus (2007d)) then we find a relatively high coefficient of correlation, 67%. We are led to think that colour or shop might be a good indicator of a fit.

```
NominalCorrelation[mat] // N
```

0.665851

Above contingency table can be collapsed when we join the two shops:

```
matsum = Plus @@ mat
```

$$\begin{pmatrix} 7 & 9 \\ 9 & 7 \end{pmatrix}$$

When the tables only concern the problem of fitting hats then it makes sense to add the two tables, since this eliminates the confounder, i.e. the dispersion over the shops. Then we find a much smaller (and negative) association.

```
NominalCorrelation[matsum] // N
```

−0.125

The paradoxical aspect in this case can be shown by using the odds to express whether Green hats or Blue hats fit better. In Shop1 the Green hats have a "Fit / No fit" odds of 5 to 1 and the Blue hats have an odds of 8 to 2 (or 4 to 1). Hence the odds ratio is (5 / 1) / (4 / 1) = 5 / 4. We find two results: (i) for the two shops separately the odds ratio is larger than 1 (preferring Green) but (ii) for the total (7/9 versus 9/7) it is below 1 (preferring Blue).

```
OddsRatio /@ Append[mat, matsum]
```

$$\left\{\frac{5}{4}, \frac{5}{4}, \frac{49}{81}\right\}$$

From this example it transpires that we have more instruments available to say something about the association of variables. Concepts like for example correlation and odds help to clarify the structure of the data. They not only have a role next to statistical testing but they may be better suitable for cases like *confounding*.

The Simpson paradox clarifies that *confounding* is one aspect in the Lucia de Berk court case.

NB 1. These data tables might represent another kind of problem, alternatively, in which it is not sensible to merely add the tables over the shop dimension. We might do a meta-analysis on the findings of the separate shops, aggregating the problem in such a way that the overall direction reflects the individual ones. Such paradoxical results for example also happen in voting theory e.g. when there can be different districts (Saari (2001)).

NB 2. To base a decision on a coefficient of correlation, one might specify a confidence interval, as can be done for real data and ordinary least squares regression, see Colignatus (2006). Such an interval has not yet been developed for this measure for nominal data. However, commonly one will regard -0.125 as a low correlation and 0.67 as a higher correlation, and one may take this as indicative. See Colignatus (2007d, f, g) for a discussion of nominal correlation. The application of this theory to the Lucia de Berk case would be a first application so that it would need to be treated with special care. The crux of the argument for the Lucia de Berk case lies in standard statistical insights but the insights from nominal correlation highlight those.

# The "data" and their correction

### ■ In general

Originally, Elffers (2002) used "data" that later were shown to be a result of data-selection. Derksen (2006) provides various corrections, for example also extending the sample period. We just consider a part of those corrections. We will use the labels "Original" and "Derksen" to identify the two sets of numbers.

We will consider two hospitals, JKZ and RKZ, and two departments in the latter, RKZ1 and RKZ2. We count the number of roster-services and the number of "incidents". There is a nurse called *V* and there are the other nurses.

### ■ The original "data"

Note that the "Original" data of the JKZ are very ominous for *V* since all "incidents" only occur during her shift. This is not the case in the RKZ. This is an indication that there may have been selection of the data (but Elffers (2002) takes the "data" as they "are").

```
TableForm[Data[Original], TableHeadings → Data[TableHeadings, All] ]
```

|      |       | Incident | No incident | Sum  |
|------|-------|----------|-------------|------|
| JKZ  | V     | 8        | 134         | 142  |
|      | Other | 0        | 887         | 887  |
|      | Total | 8        | 1021        | 1029 |
| RKZ1 | V     | 1        | 0           | 1    |
|      | Other | 4        | 361         | 365  |
|      | Total | 5        | 361         | 366  |
| RKZ2 | V     | 5        | 53          | 58   |
|      | Other | 9        | 272         | 281  |
|      | Total | 14       | 325         | 339  |
| All  | V     | 14       | 187         | 201  |
|      | Other | 13       | 1520        | 1533 |
|      | Total | 27       | 1707        | 1734 |

We find a correlation of 33.7%, indicating that there is some relationship within these data.

```
result[Original, "Not summed"] =
 {Correlation → NominalCorrelation[Data[Original, Inner]] // N}
```

{Correlation → 0.337002}

However, the dispersion over the different hospitals and departments may well be a confounder. Since *V* changed job stations it would be fair to compare her with someone who had stayed put. This is actually a dispersion over time and time is known to be an important confounder. Indeed, adding over the hospital departments, we find only a correlation of 15.8%.

    **Data[Original, Sum, Inner]**

$\begin{pmatrix} 14 & 187 \\ 13 & 1520 \end{pmatrix}$

    **result[Original, "Summed"] =
    {Correlation → NominalCorrelation[Data[Original, Sum, Inner]] // N}**

{Correlation → 0.158169}

The measure of correlation for nominal data means that it expresses the ratio of the area of the parallelepiped, created by the data vectors, with the total area of rectangle, created by the row and/or column sums. The number of 15.8% also includes the transpose of the matrix. The following picture arises when we only consider the row vectors.

    **ShowDet[Data[Original, Sum, Inner]];**

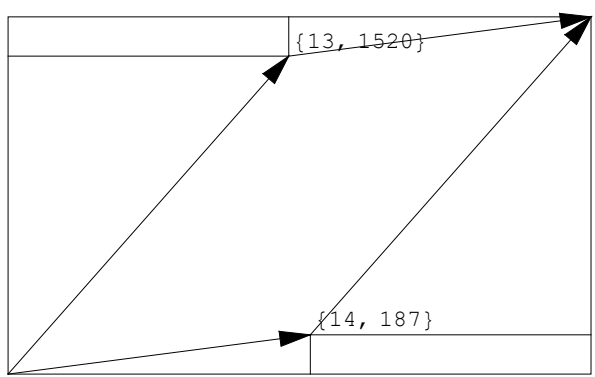

Clearly, *V* had a little more than 10% of the shifts but still the same number of "incidents" so one would tend to accept that there may have been some cause for closer investigation. (Yet, all medical reports indicated natural causes in case when there was a death, so those were only revised in the course towards the trial and during the trial).

## ■ The corrected data

Derksen (2006) gives various corrections. One of those is a conservative correction, keeping the main border totals, but checking whether the allocations fitted the rosters. Some "incidents" could be removed since those were also removed by the Court (without a recalculation of the odds).

    **TableForm[Data[Derksen], TableHeadings → Data[TableHeadings, All] ]**

|       |       | Incident | No incident | Sum  |
|-------|-------|----------|-------------|------|
|       | *V*   | 4        | 138         | 142  |
| JKZ   | Other | 1        | 886         | 887  |
|       | Total | 5        | 1024        | 1029 |
|       | *V*   | 1        | 2           | 3    |
| RKZ1  | Other | 4        | 359         | 363  |
|       | Total | 5        | 361         | 366  |
|       | *V*   | 1        | 57          | 58   |
| RKZ2  | Other | 9        | 272         | 281  |
|       | Total | 10       | 329         | 339  |
|       | *V*   | 6        | 197         | 203  |
| All   | Other | 14       | 1517        | 1531 |
|       | Total | 20       | 1714        | 1734 |

Using the corrected data, we find an association of 24.6%.

```
result[Derksen, "Not summed"] =
 {Correlation → NominalCorrelation[Data[Derksen, Inner]] // N}
```

{Correlation → 0.246024}

However, adding the hospitals and eliminating that confounder, we find a correlation of 6.1%.

```
Data[Derksen, Sum, Inner]
```

$$\begin{pmatrix} 6 & 197 \\ 14 & 1517 \end{pmatrix}$$

```
result[Derksen, "Summed"] =
 {Correlation → NominalCorrelation[Data[Derksen, Sum, Inner]] // N}
```

{Correlation → 0.0614621}

The parallellogram now takes a smaller fraction of the total area.

```
ShowDet[Data[Derksen, Sum, Inner]];
```

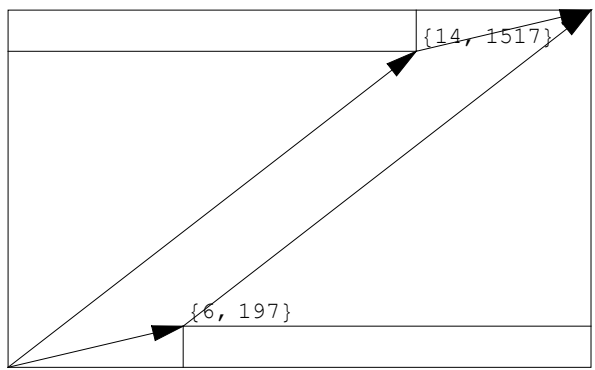

## ◼ Overview

The following table collects above results in a summary overview.

```
heading = TableHeadings → {{Original, Derksen}, {"Not summed", "Summed"}};

InsideTable[Set, result, heading]

InsideTable[Show, Correlation]
```

|  | Not summed | Summed |
|---|---|---|
| Original | 0.337002 | 0.158169 |
| Derksen | 0.246024 | 0.0614621 |

The main change is from 33.7% to 6.1%, where misleading correlation has been removed by correcting the "data" and by properly summing out. Two intermediate conclusions are possible: (1) The court case is an instance of confounding, (2) The court case is an instance of data-selection.

Another point is that the coefficient of correlation is an easy help to clarify these conclusions. In itself, correlation can be a dangerous tool. But when applied wisely it can be more transparent than other methods. An advantage is that we don't have to specify a statistical model. We can clarify that latter point by considering the original method by Elffers (2002). It appears that formulating a specific statistical model may petrify the view on the problem so that one can lose the overview.

# Using Fisher's exact (hypergeometric) method

The following doesn't cite from the official report that still is confidential, but refers to later public sources. My knowledge of the official report merely allows to make a wise selection from those public sources.

The original approach by Elffers (2002) is "on the fly". He calculates separate p-values per department using Fisher's exact method, multiplies those values, and multiplies the whole by 27, which is the official number of nurses on the roster. The latter multiplication he calls a "post-hoc correction", favourable for De Berk, given that some of the "incidents" in the JKZ caused the hypothesis of association and "thus" cannot be used for testing the hypothesis. The final number is called a "p-value" again and used in a "test", rejecting the (new) hypothesis of independence in favour of a conclusion of some relationship. Elffers mentions that the cause need not be murder and states some five alternatives, but not those of data-selection and confounding. A key point is that he explicitly excludes mere chance as a possible cause, though a statistical test still allows chance.

Elffers's original method and some alternatives are presented in Appendix A. These alternatives are: summing the "original data" to get rid of confounding, using Derksen's corrected data, and summing the latter. The following table collects the results. The table gives the 1 in N (number of nurses) probability that *V's* data would be reproduced. The "original data" with data selection and confounding suggest a condemning low probability of 1 in 342 million nurses having the same pattern while the corrected data with elimination of confounding by summing the data suggest that 1 in 1.6 nurses have that same pattern.

|          | Not summed           | Summed   |
|----------|----------------------|----------|
| Original | $3.42638 \times 10^8$ | 141494.  |
| Derksen  | 688.367              | 1.64051  |

The conclusions of confounding and data-selection can be seen from this table too. Yet in this case one requires a model. The current model uses the hypergeometric assumption (no replacement), the assumption of "at least" versus "exactly the same", multiplication of probabilities, the "post-hoc correction". And one would need a significance level to decide the issue. Perhaps the numbers in the "summed" column are sufficiently small and the numbers in the "not summed" column are sufficiently high to discard the choice of a significance level. Yet, overall, the correlation measure seems more straightforward in that it doesn't require these assumptions and by itself clearly shows the statistical mistakes in this court case.

A major point is that using the model of the hypergeometric distribution tends to focus attention on the lowest aggregation level so that one may forget to aggregate the data to eliminate confounding. Hence, in general it would seem to be wise to check upon data selection and confounding before plunging into a specific model.

PM. It is useful to express doubts about the "post-hoc correction". Let us consider a general statistical question. In a first stage of investigation the data may be collected in a makeshift fasion and then cause the formation of a hypothesis. After that, those data are less useful because of their makeshift fashion. The second stage would be a randomized control study, with a proper data collection protocol, and taking care that randomization does not itself confound the issue (e.g. drawing risk-takers). It is this kind of thinking that induces us to say that we cannot use the data that caused the hypothesis to test the hypothesis. However, if the data are merely used in a prosecution phase to identify potential questionable incidents that require closer investigation, then the issue turns into an issue of controlling the cost of investigation and medical re-examination, and then it wouldn't matter whether one includes the data that caused the hypothesis, since all cases involved are all roughly defined "incidents" in the first place anyway. The only decision that one would take would be whether the "incidents" would require closer medical examination or not. This is an entirely different issue than deciding on murder and guilt. Thus, instead of applying a "post-hoc correction", Elffers (2002) should have restricted his conclusions to the prosecution phase of identifying cases for closer investigation. Instead of expressing that chance was "excluded" (contributing to the idea that there was a serial killer) he rather should have expressed that one could accept the cost of medically re-examining the reported (and non-reported) "incidents".

## Some caveats

### ■ In general

This paper only reviewed some of the basic statistical issues of this case. The statistical analysis can be extended with heterogeneity of probabilities across nurses, a decomposition in tenured and temporary nurses, time-series / survival analysis, epidemiology using a national reference, a categorization of "incidences" by qualities such as day and night shift, and so on. Unfortunately, such basic data are mostly lacking. One of the points of critique with respect to Elffers (2002) indeed is that there is no elementary report on the data.

### ■ Heterogeneity

Heterogeneity means that nurses may have different rates of occurrence of "incidents". Some may have more, some less, with possibly a systematic difference relating to the kind or type of nurse or person. The real issue is not whether *V* was an "average nurse" but whether her record falls into what can happen amongst nurses. The model would generate a heterogeneous probability $p_i$ of an "incidence per shift for nurse *i* ". Of the JKZ we know that there were 26 other nurses (see that "post hoc correction"), but we don't have their individual scores. For the RKZ we don't know even their total number. Thus we cannot pinpoint that heterogeneity. But we can see a bit of it.

For the average $p_0$ of "incidences per shift per average nurse" the total number *n* of non-*V* nurses drops from the calculation (assuming that we take those others as the standard, which one might question).

```
p0 = (13 / n) / (1533. / n)
```

0.0084801

We see variation when we do the same kind of calculation in JKZ, RKZ1 and RKZ2 for the other nurses:

```
Module[{a}, a = Data[Original, #][[2]]; a[[1]] / a[[3]] // N] & /@ {JKZ, RKZ1, RKZ2}
```

{0., 0.0109589, 0.0320285}

For the various combinations we get the following table (calculation not shown):

```
heading2 = TableHeadings → {{Original, Derksen}, {"V", "Other"}};

InsideTable[Set, probav, heading2]
```

**InsideTable[Show, Pr]**

|          | V         | Other     |
|----------|-----------|-----------|
|          | 0.056338  | 0.        |
| Original | 1.        | 0.0109589 |
|          | 0.0862069 | 0.0320285 |
|          | 0.028169  | 0.0011274 |
| Derksen  | 0.333333  | 0.0110193 |
|          | 0.0172414 | 0.0320285 |

These calculations all assume an average nurse (per ward) and thus hide individual heterogeneity. Of course, the number of observations per ward is smaller now, so that it may be difficult to distinguish between the underlying heterogeneity and the probability resulting from drawing from these distributions. Yet, if such heterogeneity exists then the model should account for it.

### ■ The binomial model

When we consider the outcome for a single nurse, *V* might be the very example.

*V* has a higher relative frequency $p_1$ compared to $p_0$. Is this a "personal probability" or can this be seen as a cumulative probability resulting from drawing $n = 201$ times from that distribution with probability $p_0$ ? In the original "data" the heterogeneity is amplified by data-selection.

**p1 = 14 / 201.**

0.0696517

Proceeding along this line gives us the binomial model. This approach is put in Appendix B below, and its results have also been mentioned in Colignatus (2007c). Since the model uses explicit $p_i$ it might be easier to grasp than the hypergeometric model. In both models we consider each shift as a draw and thus we still do not allow that there might be more "incidents" in a shift. The hypergeometric model expresses drawing without replacement. Each "incident" reduces the probability of an "incident" in a next draw. The binomial model expresses drawing with replacement. In a simulation, we would keep the number of draws the same but the number of "incidents" could differ per simulation. The difference thus is that the hypergeometric model has a fixed target of total "successes" while the binomial model allows some variation in that total too.

In the hypergeometric model, when an "incident" has occurred, it precludes another occurrence, while in the binomial model an "incident" doesn't stop another "incident" from occurring. It is important to observe that both the hypergeometric model that Elffers (2002) used in Appendix A and the binomial model in Appendix B both assume a homogeneity over shifts. They exclude a heterogeneity between nurses and their shifts. The distinction between $p_0$ and $p_1$ already introduces the possibility of some heterogeneity, but mainly expresses that *V* is not a non-*V,* while real heterogeneity would allow all $p_i$ to differ. In fact, that latter approach seems most relevant here. But there are no data to develop that (except making additional assumptions on the same data).

## Conclusion

The objective of this paper was to clarify the issues of data-selection and confounding in the court case of Lucia de Berk. The use of the concept of "nominal correlation" allows us to give this clarification without relying on the notion of a statistical test. Reviewing the discussion you might agree that this presentation indeed has contributed to clarity.

Some of the conclusions in Colignatus (2007c) can be repeated here. A first advice is that the use of statistics is restricted, as much as possible, to the prosecution preparation stage and that its stays, as much as possible, out of the Court room. Statistics could help to identify issues and events that need closer investigation by the prosecution, but decisions about acquittal or further prosecution in Court should better be based on proper evidence from the crime scene only - if the term "crime scene" is proper anyway when there is no crime to start with. This advice merely repeats what has been stated by many others before. In fact, the Dutch Penal Court of Appeal in 2004 decided to drop the statistical calculations by Elffers (2002). A problem however is that the spirit of those calculations can still be observed in that same verdict on appeal. It still exhibits the notion that these "incidences" couldn't be a chance occurrence, while the verdict still allows "chaining". This, plus the observation by Derksen (2006) that there aren't even 2 "cases" to start with, are important considerations to reopen the court case. The exclusion of statistics can also be an error since the prosecution shouldn't exclude proper statistical calculations since these would actually form a ground to acquit Lucia de Berk from prosecution (at least on the "chained cases") given the data-selection and confounding.

# Appendix A on using Fisher's exact (hypergeometric) method

### ■ The original approach by Elffers (2002)

The following doesn't cite from the official report that still is confidential, but refers to later public sources. My knowledge of the official report merely allows to make a wise selection from those public sources.

The original approach by Elffers (2002) is "on the fly". He calculates separate p-values per department using Fisher's exact method, multiplies those values, and multiplies the whole by 27, which is the official number of nurses on the roster. The latter multiplication he calls a "post-hoc correction", favourable for De Berk, given that some of the "incidents" in the JKZ caused the hypothesis of association and "thus" cannot be used for testing the hypothesis. The final number is called a "p-value" again and used in a "test", rejecting the (new) hypothesis of independence in favour of a conclusion of some relationship. Elffers mentions that the cause need not be murder and states some five alternatives, but not those of data-selection and confounding. A key point is that he explicitly excludes mere chance as a possible cause, though a statistical test still allows chance.

The hypergeometric distribution models the case when one draws without replacement from an urn with known numbers of green and blue balls.

$n$ = total number of shifts
$r$ = Lucia's shifts
$k$ = incidents
$x$ = Lucia's incidents

```
PDF[HypergeometricDistribution[r, k, n], x]
```

$$\frac{\binom{k}{x}\binom{n-k}{r-x}}{\binom{n}{r}}$$

Let us consider the RKZ2 department where *V* had 5 incidents while the total was 14. Let us consider the probability that *V* had *at least* the number that she scored. Allowing that *V* might have had 5 to 14 incidents means that the probability is larger, which is favourable for *V*. (Note that a low probability would mean a rejection of the hypothesis of independence.)

> **Data[Original, RKZ2]**

$$\begin{pmatrix} 5 & 53 & 58 \\ 9 & 272 & 281 \\ 14 & 325 & 339 \end{pmatrix}$$

> **Sum[PDF[HypergeometricDistribution[58, 14, 339], x], {x, 5, 14}] // N**

0.0715592

The routine FisherExactPr calculates this probability for any 2 × 2 table (with or without border sums).

Doing the calculations for the three departments:

> **p[JKZ] = FisherExactPr[Data[Original, JKZ]] // N**

$1.10572 \times 10^{-7}$

> **p[RKZ1] = FisherExactPr[Data[Original, RKZ1]] // N**

0.0136612

> **p[RKZ2] = FisherExactPr[Data[Original, RKZ2]] // N**

0.0715592

> **pElffers = 27 p[JKZ] p[RKZ1] p[RKZ2]**

$2.91853 \times 10^{-9}$

And the number of 1 in 342 million, mentioned in Court and the newsmedia, can be found by dividing the latter.

> **OneInN[Original, "Not summed"] = {N → 1 / pElffers}**

$\{N \to 3.42638 \times 10^8\}$

In statistical testing, the significance level $\alpha$ must be very small to reject the hypothesis of independence.

## ■ The critique by Derksen (2006)

As stated above, Derksen (2006) gave the critique of data-selection and confounding. We add the hospitals and use the corrected data.

### ■ With the original "data"

First, we eliminate confounding, using the original "data".

**Data[Original, Sum]**

$$\begin{pmatrix} 14 & 187 & 201 \\ 13 & 1520 & 1533 \\ 27 & 1707 & 1734 \end{pmatrix}$$

**FisherExactPr[%] // N**

$2.61756 \times 10^{-7}$

Eliminating this confounder, and still using the "post-hoc correction", we find only 1 in 141494 nurses to have that pattern.

**OneInN[Original, "Summed"] = {N → 1 / (27 * %)}**

$\{N \to 141494.\}$

### ◼ With the corrected data

There was data-selection, so let us redo all calculations using the corrected data. First for all departments separately.

**(FisherExactPr[Data[Derksen, #]] // N) & /@ {JKZ, RKZ1, RKZ2}**

{0.00155956, 0.0405357, 0.851093}

**pDerksen = 27 * Times @@ %**

0.00145271

**OneInN[Derksen, "Not summed"] = {N → 1 / pDerksen}**

$\{N \to 688.367\}$

Secondly when we add over the departments to eliminate the confounding dispersion.

**Data[Derksen, Sum]**

$$\begin{pmatrix} 6 & 197 & 203 \\ 14 & 1517 & 1531 \\ 20 & 1714 & 1734 \end{pmatrix}$$

**FisherExactPr[%] // N**

0.0225766

**OneInN[Derksen, "Summed"] = {N → 1 / (27 %)}**

$\{N \to 1.64051\}$

Thus 1 in 1.6 nurses might have experienced the same (though likely not with the same chance of subsequent data-selection and confounding).

### ◾ Overview

The following table collects above results in a summary overview. The table gives the 1 in N (number of nurses) probability that *V's* data would be reproduced.

**InsideTable[Set, OneInN, heading]**

**InsideTable[Show, N]**

|          | Not summed           | Summed  |
|----------|----------------------|---------|
| Original | $3.42638 \times 10^8$ | 141494. |
| Derksen  | 688.367              | 1.64051 |

See the further discussion in the main body of the text.

## Appendix B on the binomial model

### ◾ In general

The application of this model again highlights the issues of confounding and data selection. This Appendix need not be mentioned for this paper, but has been included here anyway since its results have been used in Colignatus (2007c) and it is useful to document it.

### ◾ The binomial with the original "data"

Recall the original "data".

**Data[Original, Sum]**

$$\begin{pmatrix} 14 & 187 & 201 \\ 13 & 1520 & 1533 \\ 27 & 1707 & 1734 \end{pmatrix}$$

If *V* draws *n* = 201 times with replacement from the non-*V* distribution, then we expect the following distribution (density):

**PDF[BinomialDistribution[201, 13/1533.], x]**

$$0.0084801^x \, 0.99152^{201-x} \binom{201}{x}$$

From this distribution we can determine the probability that *V* has at least *k* cases (namely 1 minus the probability that *V* has 0 to *k*-1 cases). This invokes the Cumulative Distribution Function.

```
BinomialCDF[201, 13/1533., {2, 14}, -1, "Cases ≥ "]
```

$$\begin{pmatrix} \text{Cases} \geq 3 & 0.24363 \\ \text{Cases} \geq 4 & 0.0930338 \\ \text{Cases} \geq 5 & 0.0292779 \\ \text{Cases} \geq 6 & 0.00779387 \\ \text{Cases} \geq 7 & 0.00179153 \\ \text{Cases} \geq 8 & 0.00036146 \\ \text{Cases} \geq 9 & 0.0000648622 \\ \text{Cases} \geq 10 & 0.0000104641 \\ \text{Cases} \geq 11 & 1.5314 \times 10^{-6} \\ \text{Cases} \geq 12 & 2.04843 \times 10^{-7} \\ \text{Cases} \geq 13 & 2.52053 \times 10^{-8} \\ \text{Cases} \geq 14 & 2.86883 \times 10^{-9} \\ \text{Cases} \geq 15 & 3.03491 \times 10^{-10} \end{pmatrix}$$

If $V$'s relative frequency comes about from the same $p_0$ then we would expect between 4 or 5 cases. But there are 14, suggesting that she really has a higher probability of meeting with an "incident". We still wouldn't know what the cause is, and it might still be some natural heterogeneity amongst nurses.

Given that there are at least 14 cases, one would regard $V$ as a non-normal nurse, with a 1 in 348 million probability that this is a wrong view (since that is the probability that a non-$V$ nurse has at least 14 "incidents").

```
%[[-2]]
```

{Cases ≥ 14, $2.86883 \times 10^{-9}$}

```
%[[2]]
```

$2.86883 \times 10^{-9}$

```
1 / %
```

$3.48574 \times 10^8$

Note that this still assumes that these "data" are correct.

### ◼ The binomial with the corrected data

Again we take the summed data.

```
Data[Derksen, Sum]
```

$$\begin{pmatrix} 6 & 197 & 203 \\ 14 & 1517 & 1531 \\ 20 & 1714 & 1734 \end{pmatrix}$$

The non-$V$ nurses form a control-group again and their probability is $p_0$.

```
p0 = 14 / 1531.
```

0.00914435

As said, compared to that, *V* has a much higher relative frequency $p_1$. However, if the true probability is $p_0$ then the outcome for *V* can still be the result of drawing *n* = 203 times from that distribution. (For a model with heterogeneity, this result would fall below the average for RKZ2.)

```
p1 = 6 / 203.
```

0.0295567

Thus, if *V* draws *n* = 203 times from that non-*V* distribution, then we expect the following distribution:

```
PDF[BinomialDistribution[203, 14 / 1531.], x]
```

$0.00914435^x \, 0.990856^{203-x} \binom{203}{x}$

From this distribution we can determine the probability that *V* has at least *k* cases (namely 1 minus the probability that *V* has 0 to *k*-1 cases).

```
BinomialCDF[203, 14 / 1531., {2, 8}, -1, "Cases ≥ "]
```

$\begin{pmatrix} \text{Cases} \geq 3 & 0.284318 \\ \text{Cases} \geq 4 & 0.117044 \\ \text{Cases} \geq 5 & 0.0398576 \\ \text{Cases} \geq 6 & 0.0115067 \\ \text{Cases} \geq 7 & 0.00287253 \\ \text{Cases} \geq 8 & 0.000630018 \\ \text{Cases} \geq 9 & 0.000122978 \end{pmatrix}$

If *V*'s relative frequency comes about from the same $p_0$ then we would expect between 5 or 6 cases. There are 6, suggesting that she doesn't have a higher probability than the other nurses of meeting with an "incident".

Given that there are at least 6 cases, one might regard *V* as a non-normal nurse anyhow, with a 1 in 87 (not *million*) probability that this is a wrong view (since that is the probability that a non-*V* nurse has at least 6 "incidents").

```
%[[4]]
```

{Cases ≥ 6, 0.0115067}

```
%[[2]]
```

0.0115067

```
1 / %
```

86.9055

### ◼ Intermediate conclusion

The choice of the hypergeometric distribution by Elfffers (2002) likely was not adequate. The application of the binomial model creates some more perspective. The probabilities can be heterogeneous, which in itself is not strange. Or the data-selection created a heterogeneity that hid some homogeneity.

In all cases it holds that a statistical model by itself introduces probabilities so that all outcomes are events by chance, and thus Elffers's (2002) conclusion that chance is excluded is unwarranted.

In above discussion of the binomial model we have refrained from using the framework of hypothesis testing and selecting a level of significance in advance, since the research question is rather undefined. As these "incidents" aren't murders, one wouldn't want to convict someone for murder merely because of being on shift (or being more on shift than others at some level of significance). The Court already gave such a verdict but we shouldn't repeat that error. Not only the research question was vague originally as well but also the Court didn't specify the significance level and statistical power in advance, to reflect the costs of convicting an innocent person or releasing a serial killer (assuming that the prosecution can narrow down the question to that issue and doesn't merely apprehend someone based upon hearsay only).

Given all these unknown elements, and without applying hypothesis testing, it nevertheless transpires that data-selection and confounding featured strongly in this court case. The notion of "nominal correlation" shows this clearer than the hypergeometric and binomial models.